\newif\ifiop

\ifiop
\documentclass{jpconf}
\else
\documentclass[a4paper,11pt]{article}
\usepackage[backend=bibtex,citestyle=ieee,style=ieee]{biblatex}
\bibliography{causality_bib}
\usepackage{hyperref}
\fi

\usepackage[latin1]{inputenc}
\usepackage[english]{babel} 
\usepackage{amsfonts,amssymb,amsmath,amscd,mathtools,slashed,mathrsfs}
\usepackage[amsthm,thmmarks,amsmath]{ntheorem}
\usepackage{enumerate}

\newtheorem{thm}{Theorem}

\theoremstyle{definition}

\newcommand{\norm}[1]{\left\Vert {#1} \right\Vert}
\newcommand{\abs}[1]{\left\vert {#1} \right\vert}
\newcommand{\set}[1]{\left\{ {#1} \right\}}
\newcommand{\prt}[1]{\left( {#1} \right)}

\newcommand{\scal}[1]{\left< {#1} \right>}

\newcommand{\dv}[2]{\frac{\mathrm d{#1}}{\mathrm d{#2}}}

\newcommand{\setR}{{\mathbb R}}
\newcommand{\setC}{{\mathbb C}}

\newcommand{\Lequi}{\Longleftrightarrow}
\newcommand{\precc}{\prec\!\!\prec}
\newcommand{\A}{\mathcal{A}}

\renewcommand{\H}{\mathcal{H}}

\newcommand{\J}{\mathcal{J}}
\newcommand{\M}{\mathcal{M}}

\newcommand{\T}{\mathcal{T}}

\newcommand{\St}{\mathscr{S}}

\renewcommand{\M}{M}

\begin{document}

\title{The Lorentzian distance formula in noncommutative geometry}

\author{Nicolas Franco}

\ifiop
\address{Namur Center for Complex Systems (naXys) \& Department of Mathematics, University of Namur, rue de Bruxelles 61, 5000 Namur, Belgium}
\ead{nicolas.franco@unamur.be}
\else
\date{\footnotesize Namur Center for Complex Systems (naXys) \& Department of Mathematics,\\ University of Namur, rue de Bruxelles 61, 5000 Namur, Belgium\\
\href{mailto:nicolas.franco@unamur.be}{nicolas.franco@unamur.be}}
\maketitle
\fi

\begin{abstract}
For almost twenty years, a search for a Lorentzian version of the well-known Connes' distance formula has been undertaken. Several authors have contributed to this search, providing important milestones, and the time has now come to put those elements together in order to get a valid and functional formula. This paper presents a historical review of the construction and the proof of a Lorentzian distance formula suitable for noncommutative geometry. 
\end{abstract}

\ifiop
\else
\vspace*{0.5cm}
\fi

\section{Introduction and formulation of the Lorentzian distance}\label{sec_intro}

Connes' noncommutative geometry  \cite{Connes:1994aa,MC08}  provides at the same time a beautiful mathematical theory as well as new tools for physical models of unification theory. At a mathematical level, the topological correspondence between locally compact Hausdorff spaces and commutative \mbox{C$^*$-algebras} given by Gel'fand's theory is brought up to the level of Riemannian manifolds. The key elements are spectral triples $(\A,\H,D)$ from which, among others, information concerning the metric aspect can be recovered using the Riemannian distance formula :
\begin{equation}\label{rdist}
d_R(p,q) \ =\ \sup_{f \in \A}\set{ \ \abs{f(q)-f(p)} \ :\ \norm{[D,f]} \leq 1\ }\cdot
\end{equation}
Applications of noncommutative geometry in mathematical physics take part mainly in particle physics and quantum field theory. However the physical Lorentzian signature of spacetimes makes the use of the initial mathematical theory more problematic, especially concerning the formula \eqref{rdist}. Two paths have been followed to solve this problem: the Wick rotation process which allows the use of all the well-defined tools of Riemannian noncommutative geometry (see e.g.~\cite{Schcker2005,Wickrotation}), but therefore with a loss of causal information, or the adaptation of the theory to a Lorentzian signature, which is less straightforward and still ongoing, notably with the use of Krein spaces and Lorentzian spectral triples \cite{Stro,F5,Rennie12,BESNARD2017}. In this last context, several authors have tried to generalize the formula \eqref{rdist} to a Lorentzian distance formula \cite{ParfZap,Moretti,F2,F3,CQG2013,RENNIE2016108,Ming,Ming2}. Each of those authors has significantly contributed to a specific step of the construction of a final formula. In this paper, we will go through the summary of those different steps and we present two formulations of a now completely proved Lorentzian distance formula.

The first formulation is at the level of traditional Lorentzian geometry, where the usual Lorentzian distance $d(p,q)$ between two points, representing the maximal length of the piecewise $C^1$ future-directed causal curves from $p$ to $q$ \cite{beem}, is rewritten in a completely path-independent way, using the information coming from a specific set of test functions. Key elements of the proof of the following formula will be presented in Section \ref{sec_path}.

\ifiop
\else
\vspace*{0.2cm}
\fi

\begin{thm}[Path-independent formulation]\label{th_path-ind}\ \\
If $(\M,g)$ is a time-oriented Lorentzian manifold (spacetime) which is either:
\begin{itemize}
\item globally hyperbolic,
\item stably causal such that the usual Lorentzian distance $d$ is continuous and finite,
\end{itemize}
then for all $p,q \in \M$:
\begin{equation}\label{path-ind}
 d(p,q) \ =\ \inf\set{ \ [f(q)-f(p)]^+ \ :\ f \in \St\ },
 \end{equation}
where $[\alpha]^+ = \max\set{0,\alpha}$ and $\St$ is the set of smooth real-valued ``steep'' functions, i.e.~the set of \mbox{$f \in C^1(\M,\setR)$} such that  \mbox{$g( \nabla f, \nabla f ) = g^{-1}( df, df ) \leq -1 $} and $ \nabla f$ is past-directed ($f$ is a future-directed temporal function).
\end{thm}
\ifiop
\else
\vspace*{0.2cm}
\fi

Stable causality is the weakest assumption under which the RHS of \eqref{path-ind} makes sense, otherwise the set of steep functions $\St$ is empty \cite{MS08}. The condition of continuity of the Lorentzian distance $d$ is necessary since the RHS is upper semi-continuous while the LHS is lower semi-continuous \cite{Ming}. Under such an assumption, $(M,g)$ is in fact a causally continuous spacetime in the sense of \cite{hawking1975large}. The condition of global hyperbolicity is a particular case where the Lorentzian distance $d$ is automatically continuous and finite \cite{beem}.

The second formulation is an algebraic formulation, where every element from traditional Lorentzian geometry has been replaced by a corresponding element coming from the theory of spectral triples. This formulation opens the possibility of a generalization to noncommutative spacetimes. The proof will be presented in Section \ref{sec_ncg} while the possible technical difficulties for an application on noncommutative spacetimes will be presented in Section \ref{sec_diff}.

\ifiop
\else
\newpage
\vspace*{0.3cm}
\fi

\begin{thm}[Spectral triple formulation]\label{th_op_form}\ \\
If $(\M,g)$ is a $n$-dimensional spin Lorentzian manifold which is either globally hyperbolic or stably causal such that the Lorentzian distance $d$ is continuous and finite, and if we define:
\begin{itemize}
\item  The algebra $\A = C^1(\M,\setR)$ with pointwise multiplication,
\item The Hilbert space $\H = L^2(\M,S)$ of square integrable sections of the spinor bundle $S$ over~$\M$ (using a positive definite inner product on the spinor bundle),
\item The Dirac operator $D = -i(\hat c \circ \nabla^S) = -i e^{\mu}_a\gamma^{a} \nabla^S_\mu$ associated with the spin connection~$\nabla^S$,
\item The fundamental symmetry $\J=i\gamma^0$ where $\gamma^0$ is the first flat gamma matrix\footnote{Conventions
used in the paper are $(-,+,+,+,\cdots)$ for the signature of the metric and
$\{\gamma^a,\gamma^b\}=2\eta^{ab}$ for the flat gamma matrices, with $\gamma^0$ anti-Hermitian and
$\gamma^a$ Hermitian for $a>0$. },
\item If $n$ is even, the chirality operator $\chi = \pm i^{\frac{n}{2} +1} \gamma^0 \cdots \gamma^{n-1},$
\end{itemize}
\vspace*{0.5cm}
then for all $p,q \in \M$, if $n$ is even:
\begin{equation}\label{op_form}
 d(p,q) \ =\ \inf_{f \in \A} \set{ \ [f(q)-f(p)]^+ \ :\ \forall \phi \in \H, \scal{\phi,\J([D,f]+ i\chi) \phi } \leq 0\ },
 \end{equation}
 and if $n$ is odd:
 \begin{equation}\label{op_form2}
 d(p,q) \ =\ \inf_{f \in \A} \set{ \ [f(q)-f(p)]^+ \ :\ \forall \phi \in \H, \scal{\phi,\J([D,f]\pm 1) \phi } \leq 0\ },
 \end{equation}
where $[\alpha]^+ = \max\set{0,\alpha}$  and $\scal{\cdot,\cdot}$ is the positive definite inner product on $\H$.
\end{thm}

\ifiop
\else
\vspace*{0.5cm}
\fi

\section{Historical construction of the Riemannian and Lorentzian distance formulas}\label{sec_histo}

The first apparition of Connes' distance formula \eqref{rdist} comes from 1989 in \cite{connes_1989}. Common presentations and proofs of the formula are given by \cite{Connes1992}, \cite[Chapter~6]{Connes:1994aa} and \cite[Chapter~9.3]{Elements}. The formula has been studied and applied by many authors on several kinds of spaces, as among others \cite{BIMONTE1994139,RieffelStates,IochKM,MoyalDist1,MoyalDist2,MoyalDist3,FrancoMoyal}.

\newpage

The way to prove this formula is quite direct. If we consider a connected compact Riemannian manifold $(M,g)$ and two points $p$ and $q$ on it, we can choose an arbitrary piecewise $C^1$ curve $\gamma : [0,1] \rightarrow M$ with $\gamma(0) = p$ and $\gamma(1) = q$. Then, for each function $f \in C^\infty(M)$, we have by using the second fundamental theorem of calculus:
\begin{eqnarray*}
f(q) - f(p) &=& f(\gamma(1)) - f(\gamma(0)) = \int_0^1 \dv{}{t} f(\gamma(t)) \, dt\\
 &=& \int_0^1 df(\dot\gamma(t)) \, dt = \int_0^1  g( \nabla f,\dot\gamma(t)) \, dt.
\end{eqnarray*}
Using the Cauchy--Schwarz inequality, we get:
\begin{eqnarray}
\abs{ f(q) - f(p) } &\leq& \int_0^1  \abs{g( \nabla f,\dot\gamma(t))} \, dt \leq \int_0^1  \abs{ \nabla f} \abs{\dot\gamma(t)} \, dt \nonumber\\
&\leq& \norm{  \nabla f}_\infty  \int_0^1  \abs{\dot\gamma(t)} \, dt = \norm{  \nabla f}_\infty l(\gamma), \label{r_apply_cauchy}
\end{eqnarray}
where $l(\gamma)$ denotes the length of the curve. So we obtain the following inequality:
\begin{equation}\label{r_ineq}
d_R(p,q) \geq \sup\set{  \abs{f(q)-f(p)} \ :\  f \in \A,\  \norm{  \nabla f}_\infty \leq 1 }\cdot
\end{equation}
The condition $\norm{  \nabla f}_\infty \leq 1$ can be replaced by a weaker condition \mbox{$\text{ess} \sup \norm{  \nabla f} \leq 1$} which allows us to work with the set $\A \subset C(M)$ of Lipschitz continuous functions on $M$ \cite{WEAVER1996261}.  Within this larger set, the equality is easily given by the usual distance as a function of its second argument $f(\cdot)  = d_R(p,\cdot)$. Indeed, $d_R$ is Lipschitz continuous with $\norm{  \nabla d_{R}} = 1$ except on a set of measure zero (the point $p$ and the cut locus), and we get the path-independent formula:
\begin{equation}\label{r_eq}
d_R(p,q) = \sup\set{  \abs{f(q)-f(p)} \ :\  f \in \A,\  \text{ess} \sup \norm{  \nabla f} \leq 1 }\cdot
\end{equation}
The last step is the translation of this formula into an algebraic formalism. If $M$ is a~spin manifold and $D$ the Dirac operator, we have \cite[Chapter~9.3]{Elements}:
\begin{equation}\label{r_op}
\text{ess} \sup \norm{  \nabla f} \leq 1\ \Lequi\ \norm{[D,f]} \leq 1
\end{equation}
which gives the formula \eqref{rdist}.\\

The construction of the Riemannian distance formula can be clearly divided in three important steps: the setting of a path-independent inequality \eqref{r_ineq}, the construction of the equality case \eqref{r_eq} and the operatorial (spectral triple) formulation \eqref{r_op}. The search for a Lorentzian equivalent formula went through the same three steps and we summarize here its historical evolution:

\begin{itemize}
\item 1998-2000, G. N. Parfionov and R. R. Zapatrin \cite{ParfZap}: First mention of the duality (inversion supremum-infimum and the inequality signs) in the formula \eqref{r_ineq} in a Lorentzian context.
\item 2002-2003, V. Moretti \cite{Moretti}: Generalization of the formula \eqref{r_eq} for globally hyperbolic spacetimes using a local condition on the gradient $ \nabla f$ (in a more recent terminology: using functions that are ``steep'' almost-everywhere but only inside some specific compact sets) and an attempt at algebraization using the Laplace-Beltrami-d'Alembert operator and a net of Hilbert spaces.
\item 2010, N. Franco \cite{F3}: Generalization of the formula \eqref{r_eq} for globally hyperbolic spacetimes using a global condition on the gradient $ \nabla f$ (using functions that are steep almost-everywhere on the whole spacetime). The global behavior of the test functions is chosen in order to facilitate a future algebraization. The proof of the equality case is done using non-Lipschitz continuous causal functions.
\item 2012-2013, N. Franco and M. Eckstein \cite{CQG2013}: Algebraic formulation of the global condition on the gradient (steep) for $C^1$ functions, so a Lorentzian generalization of \eqref{r_op}. However, this algebraic formulation is not valid for non-Lipschitz continuous functions as needed for the general proof in \cite{F3}, so the proof of the distance formula is limited to spacetimes where the usual distance function can be suitably approximated by $C^1$ steep functions. A particular proof for the Minkowskian case is given.
\item 2014-2016, A. Rennie and B. E. Whale \cite{RENNIE2016108}: Extension of the formula obtained in \cite{F3} for non-globally hyperbolic spacetimes. The correspondence is extended to spacetimes where the usual Lorentzian distance is finite, while conjecturing that the condition of stable causality should be necessary if the distance is also continuous. The steep condition is also proved to be necessary for a formulation in term of test functions.
\item 2017, E. Minguzzi \cite{Ming}: As a consequence of the study of causality under less-regular differentiability with the smoothing of non-Lipschitz continuous steep functions, a $C^1$ proof of the formula given in \cite{F3} is obtained, with the necessary and sufficient condition that the spacetime must be stably causal and the usual Lorentzian distance function finite and continuous. This gives a complete smooth validation of the results presented in \cite{F3,CQG2013,RENNIE2016108}. An additional $C^1$ proof of the formula, simplest but under the assumption of a globally hyperbolic spacetimes, is also presented in \cite{Ming2}.
\end{itemize}

\section{The path-independent formulation of the Lorentzian distance}\label{sec_path}

In this Section, we summarize the main arguments of the proof of Theorem \ref{th_path-ind}. The key elements of the Lorentzian distance formula are:
\begin{itemize}
\item The real-valued (continuous or not) causal functions, which are the functions which do not decrease along every future-directed causal curve.
\item The steep functions, which are $C^1$ causal functions which increase sufficiently rapidly along every future-directed causal curve., i.e.~with \mbox{$g( \nabla f, \nabla f ) \leq -1 $} and past-directed gradient.
\end{itemize}
Unlike the Riemannian distance formula \eqref{rdist}, we have to consider real-valued functions instead of complex ones in order to reach a non-symmetric formula.

At first, we need an inequality which is a Lorentzian generalization of \eqref{r_ineq}. The same two theorems are used in their existing Lorentzian versions:
\begin{itemize}
\item The second fundamental theorem of calculus, valid for absolutely continuous, hence $C^1$, functions.
\item The reverse Cauchy-Schwarz inequality \cite{beem}:
\begin{equation*}
\text{If $v$ and $w$ are timelike vectors, then } \abs{g(v,w)} \geq \sqrt{- g(v,v)} \sqrt{- g(w,w)}.
\end{equation*}
\end{itemize}

Now we consider a time-oriented Lorentzian manifold $(M,g)$ and two points $p$ and $q$ on it such that $p \precc q$. We can choose a  piecewise $C^1$ future-directed timelike curve $\gamma : [0,1] \rightarrow M$ with $\gamma(0) = p$ and $\gamma(1) = q$. Then, for each function $f \in C^1(M,\setR)$, 
\begin{equation}\label{eq_abscont}
f(q) - f(p) = \int_0^1 \dv{}{t} f(\gamma(t)) \, dt =  \int_0^1  g( \nabla f,\dot\gamma(t)) \, dt.
\end{equation}
Since $\dot\gamma(t)$ is almost everywhere a future-directed timelike vector, if we assume that $\nabla f$ is everywhere timelike with constant past-directed orientation, then the sign of $g( \nabla f,\dot\gamma(t))$ is constant and positive, so we get:
\ifiop
\else
\vspace*{-0.6cm}
\fi

\begin{eqnarray}
f(q) - f(p) &=& \int_0^1  g( \nabla f,\dot\gamma(t)) \, dt = \int_0^1 \abs{ g( \nabla f,\dot\gamma(t)) }\, dt \nonumber\\
& \geq& \int_0^1  \sqrt{-g(\nabla f,\nabla f)} \sqrt{-g(\dot\gamma(t),\dot\gamma(t))} \, dt \nonumber\\
&\geq&  \inf\set{{\sqrt{-g(\nabla f,\nabla f)}}}\; l(\gamma), \label{eq_abscont2}
\end{eqnarray}
which is the Lorentzian counterpart of \eqref{r_apply_cauchy}. This result can be extended by continuity to future-directed causal curves and $p \preceq q$. Then taking the supremum over all future-directed causal curves from $p$ to $q$ we get the following path-independent inequality:
\begin{equation}\label{c1_ineq}
 d(p,q) \ \leq\ \inf_{f \in C^1(M,\setR)}\set{ \ [f(q)-f(p)]^+ \ : \begin{subarray}{c} g( \nabla f, \nabla f ) \leq -1,\\ \ \nabla f \text{ past-directed}\end{subarray}}\cdot
 \end{equation}
 Note that this result can be extended to non-absolutely continuous functions (hence non-Lipschitz continuous) as long as we impose that the functions $f$ remain causal. Indeed, from their monotony, those functions are a.e.~differentiable on any future-directed causal curve $\gamma : [0,1] \rightarrow M$ and instead of \eqref{eq_abscont} we have the inequality $f(\gamma(1)) - f(\gamma(0)) \geq \int_0^1 \dv{}{t} f(\gamma(t)) \, dt$ which leads to the same formulation \eqref{eq_abscont2} and to the following formula:
 \begin{equation}\label{causal_ineq}
 d(p,q) \ \leq\ \inf_{\begin{subarray}{c}\text{causal}\\ \text{functions }f\end{subarray}}\set{ \ [f(q)-f(p)]^+ \ : \begin{subarray}{c}\text{\rm ess} \sup g( \nabla f, \nabla f ) \leq -1,\\ \ \nabla f \text{ past-directed}\end{subarray}}\cdot
 \end{equation}
  Three proofs have been given concerning the equality between the usual Lorentzian distance and the formulas \eqref{c1_ineq} or \eqref{causal_ineq}. Due to the length of those technical proofs, we only present the main ideas here.
  
 Under the condition of global hyperbolicity, a proof of the equality for the non-smooth formulation \eqref{causal_ineq} is presented in \cite{F3}, with the construction of a sequence of a.e.~steep continuous functions converging to the equality and constructed as locally finite sums of distance functions computed from points located near a suitable Cauchy surface. This particular proof can now be transformed in a smooth version using the new results from \cite{Ming}, which is explicitly done in \cite{Ming2}, but it is still limited to globally hyperbolic spacetimes due to the use of a Cauchy surface as main element. The second proof presented in \cite{RENNIE2016108} gets rid of the Cauchy surface and instead uses the construction of a specific achronal surface $S$ such that $\M=I^+(S)\cup S \cup I^-(S)$ and considers the distance $f(\cdot)=d(S,\cdot)$ to this surface as the equality function. This extends the proof of the non-smooth formulation \eqref{causal_ineq} to spacetimes with finite Lorentzian distance (where the distance is also allowed to be non-continuous).
 
 The most recent and general proof can be found in \cite{Ming} and concerns the smooth formulation \eqref{c1_ineq}. It implies the precedent proofs (at least when the distance is continuous). The necessary and sufficient conditions are the stable causality of the spacetime and finiteness and continuity of the Lorentzian distance (which is automatic if the spacetime is globally hyperbolic). The approach is here completely different and uses the idea that metric properties of spacetimes can be computed using a causal theory on a space with one extra dimension $\tilde M = M \times \setR$. The usual Lorentzian distance function can then be traded for a Lorentz-Finsler function defined on causal tangent vectors of the product space. The final proofs of Theorem \ref{th_path-ind} is then given by \cite[Theorem 4.11]{Ming} in the globally hyperbolic case and by \cite[Theorem 4.15]{Ming} in the more general case of stably causal spacetimes, requiring the additional condition of finiteness and continuity of the Lorentzian distance.

\section{The algebraic formulation of the Lorentzian distance}\label{sec_ncg}

In this Section, we present the proof of Theorem \ref{th_op_form} and especially the origin of the algebraic constraint inside \eqref{op_form} and \eqref{op_form2}. Once more, we will see that the metric information emerges naturally from the causal information on a space with one extra dimension. So at first we need to review the causal theory for spectral triples.

We consider a stably causal Lorentzian $n$-dimensional manifold $(\M,g)$ with a spin structure $S$ with its associated space $L^2(\M,S)$ of square integrable sections of the spinor bundle over~$\M$. This space is naturally endowed with a indefinite inner product $\prt{\cdot,\cdot}$ coming from the spin structure and possesses all the properties of a Krein space \cite{Bog}, but can be turned into a Hilbert space $\H$ using an alternative positive-definite inner product $\scal{\cdot,\cdot} = \prt{\cdot,\J\cdot}$ and $\prt{\cdot,\cdot} = \scal{\cdot,\J\cdot}$. The operator $\J$ is called a fundamental symmetry and can be constructed from the Clifford action of a timelike vector field \cite{Muller2014}. Since stable causality implies the existence of a smooth temporal function $\T$ \cite{MS08}, a (Hermitian) fundamental symmetry is easily given by $\J = ic(d\T) = i\gamma^0$, where $\gamma^0$ is the first flat gamma matrix respecting $(\gamma^0)^2=-1$ (up to a smooth conformal transformation which leaves the causality invariant). If we consider the Lorentzian Dirac operator $D = -i e^{\mu}_a\gamma^{a} \nabla^S_\mu$, where $e^{\mu}_a$ stand for the vierbeins\footnote{We consider here a ``pseudo-orthonormal'' frame coming from the timelike vector field $\partial_0=\partial_\T$ with $e^{0}_0=1$ and $e^{i}_0=e^{0}_i=0$ for $i=1,\dots,n-1$.}, then this operator is anti-symmetric for the Krein product, which is equivalent to say that $\J D$ is an anti-symmetric operator for the Hilbert space $\H$. Under the additional assumption of completeness of the manifold $\M$ under spacelike reflexion, $\J D$ is a skew-Hermitian operator on $\H$ \cite{Stro}.

Then we have the following theorem coming from \cite{CQG2013} :
\begin{thm}\label{th_causal}
A function $f \in C^1(\M,\setR)$ is causal if and only if
 \begin{equation*}
 \forall \phi \in \H, \scal{\phi,\J[D,f] \phi } \leq 0
 \end{equation*}
 where $\scal{\cdot,\cdot}$ is the positive definite inner product on $\H$.
\end{thm}

Two elements are important for the proof of this Theorem. At first, the absolute continuity of the function $f$. If the function is absolutely continuous, then the causal property can be fully characterized by the following conditions on its gradient:
 \begin{equation}\label{conditions}
g(\nabla f, \nabla f) = g^{\mu\nu} f_{,\mu} f_{,\nu} \leq 0, \qquad g(\nabla f,\nabla \T)  = g^{\mu0} f_{,\mu} = - f_{,0}\leq 0,  
 \end{equation}
where $df=f_{,\mu} dx^\mu$ and $x^0=\T$ is orthonormal to the others chosen local coordinates. The second element is the $C^1$ behavior. Indeed, from the continuity of the derivative, if \eqref{conditions} is false at some point of $\M$, then it must be false on some neighborhood and this information can be caught by a suitable specific spinor. From this observation, we can see that the smooth formulation of the Lorentzian distance equation \eqref{path-ind} is important for the algebraic generalization.

From the well-known property $[D,f]=-i\,c(df)$ \cite{Elements}, the proof of Theorem \ref{th_causal} relies on the fact that the matrix:
\begin{equation*}
\J[D,f]   = i\gamma^0\,(-i)\,(\gamma^a e^{\mu}_a f_{,\mu}) = \gamma^0\gamma^a e^{\mu}_a f_{,\mu}
\end{equation*}
is pointwise negative semi-definite if and only if \eqref{conditions} is respected. A first and technical proof of this equivalence is given in \cite{CQG2013}  using the technique of the characteristic polynomial.\footnote{The initial proof was done under the hypothesis of global hyperbolicity but can be extended to simply causal spacetimes by considering the pseudo-orthonormal frame.} We present here another shorter consideration suggested in \cite{2roads}. We have 
\begin{equation*}
\J[D,f]  = -f_{,0} + \gamma^0\gamma^i e^{\mu}_i f_{,\mu} = -f_{,0} + b
\end{equation*}
 for $i=1,\dots,n-1$ where the second term $b$ is Hermitian and respects $b^2=\norm{\gamma^i \gamma^j f_{,i} f_{,j}}^2$. So the spectrum of $b$ must be $\set{\pm \norm{g^{ij} f_{,i} f_{,j}}}$. Since the reduced metric $g^{ij}$ is positive definite we get that $\J[D,f] = -f_{,0} + b$ is negative semi-definite if and only if $-f_{,0} \leq 0$ and $g^{ij} f_{,i} f_{,j} \leq f^2_{,0}$, hence \eqref{conditions}.

Theorem \ref{th_causal} is very important in noncommutative geometry since its provides a way to completely characterize causality at an algebraic level, since the complete set of causal functions is sufficient to characterize the causal relations for stably causal spacetimes \cite{Ming,Bes09}. Consequences of Theorem \ref{th_causal} on noncommutative spacetimes (almost commutative spacetimes, Moyal spacetime) have already been handled with success \cite{FrancoMoyal,SIGMA2014,CC2014,JGP2015,PROC2015,Zitter}.\\

From Theorem \ref{th_causal}, we can now easily get the characterization of the steep functions used in \eqref{path-ind}. Once more, we consider a product space $\tilde M = M \times \setR$ on which we extend the Lorentzian metric $g$ to $\tilde g$ by adding $\tilde g^{nn}=1$ and $\tilde g^{\mu n}=\tilde g^{n \mu}=0$ for $\mu=0,\dots,n-1$. We will use the extended indices $\tilde a,\tilde \mu = 0,\dots, n$. We also have to extend the spin structure to the new dimension, giving an extended Dirac operator $\tilde D$. When $n$ is even, this can be done very easily by considering the chirality operator as an additional gamma matrix $\gamma^n = \pm\chi = \pm i^{\frac{n}{2} +1} \gamma^0 \cdots \gamma^{n-1}$ (with $e^{\mu}_n = e^{n}_a=0$ and $e^{n}_n=1$). Now we can consider all functions of the form $\tilde f = f - x_n \in C^1(\tilde M, \setR)$ where $f \in C^1(M, \setR)$, which trivially gives $\tilde f_{,\mu} = f_{,\mu}$ and $\tilde f_{,n} = -1$.

Then applying Theorem \ref{th_causal} to $\tilde f$ gives:
\begin{eqnarray*}
 \forall \phi \in \H, &&\scal{\phi,\J[\tilde D,\tilde f] \phi }\\ &=&  \scal{\phi, \J \prt{ -i\gamma^{\tilde a} e^{\tilde\mu}_{\tilde a} f_{,\tilde\mu} } \phi }\\
&=&  \scal{\phi,\J\prt{ -i\gamma^{a} e^{\mu}_{a} f_{,\mu} -i \gamma^n \tilde f_{,n}}  \phi } \\
&=& \scal{\phi,\J\prt{[ D, f] \pm i \chi}\phi } \ \leq\ 0
\end{eqnarray*}
which is equivalent to the fact that $\tilde f$ is causal on $(\tilde \M, \tilde g)$, i.e. $- \tilde f_{,0}=- f_{,0} \leq 0$ and:
\begin{eqnarray*}
\tilde g(\tilde\nabla \tilde f,\tilde \nabla \tilde f) = \tilde g^{\tilde\mu\tilde\nu} \tilde f_{,\tilde\mu} \tilde f_{,\tilde\nu} = g^{\mu\nu} f_{,\mu} f_{,\nu} +\tilde  g^{nn} \tilde f_{,n}\tilde f_{,n} = g(\nabla f, \nabla f) +1  \leq 0
\end{eqnarray*}
which is the exact characterization of a steep function.
The choice of $+i\chi$ or $-i\chi$ is completely arbitrary and has no influence on the formula, so from \eqref{path-ind} we can write:
\begin{equation*}
 d(p,q) \ =\ \inf_{f \in C^1(\M,\setR)} \set{ \ [f(q)-f(p)]^+ \ :\ \forall \phi \in \H, \scal{\phi,\J([D,f]+ i\chi) \phi } \leq 0\ },
 \end{equation*}
 which is valid for manifolds with even dimension $n$ respecting the conditions of Theorem \ref{th_path-ind}. So the proof of 
 the formula \eqref{op_form} is complete.
 
This process can also be applied in order to get a valid formula for odd-dimensional manifolds, but this requires a doubling of the Hilbert space $\tilde \H = \H \otimes \setC^2$ and new gamma matrices $\tilde \gamma^\mu = \gamma^\mu \otimes \sigma^1$ for $\mu=0,\dots,n-1$ and $\tilde \gamma^n = 1 \otimes \sigma^2$ where $\sigma^i$ are the Pauli matrices. The fundamental symmetry becomes $\tilde\J = \J \otimes \sigma^1$. The negative semi-definite operator becomes:
\begin{eqnarray*}
\tilde \J[\tilde D,\tilde f] &=& \prt{\J \otimes \sigma^1} \prt{ -ie^{\mu}_{a} f_{,\mu} (\gamma^{a} \otimes \sigma^1) +i (1 \otimes \sigma^2)} \\
&=& \J [D,f] \otimes 1 + \J \otimes \sigma^3.
\end{eqnarray*}
However, $\sigma^3 = \left(\begin{smallmatrix} 1&0\\ 0&-1 \end{smallmatrix}\right)$ is diagonal, which means that the constraint:
\begin{equation*}
 \forall \tilde\phi = (\phi_+,\phi_-) \in \tilde\H,\ \scal{\tilde\phi,\tilde\J[\tilde D,\tilde f] \tilde\phi } \leq 0
 \end{equation*} 
 splits into two inequalities: 
 \begin{equation*}
 \forall \phi_\pm \in \H,\ \scal{\phi_\pm,\J([D,f]\pm 1) \phi_\pm } \leq 0.
  \end{equation*}
This gives rise to formula \eqref{op_form2} and completes the proof of Theorem \ref{th_op_form}.
 
 \ifiop
\else
\vspace{1.2cm}
\fi
 
\section{Application on noncommutative spacetimes}\label{sec_diff}

Theorem \ref{th_op_form} opens the possibility to compute metrical information on noncommutative spacetimes. The formalism is the one of Lorentzian spectral triples $(\H,\A,D)$ with a given fundamental symmetry $\J$ with $\J^2=1$, $\J^*=\J$, $[\J,a]=0$, $\forall a\in{\A}$. The conditions on the operator $D$ are $\forall a\in{\A}$,  $[D,a]$ is bounded,  $a(1 +  \frac {(D D^* + D^*D)}2)^{-\frac 12}$ is compact and $D^*=-\J D \J$ (Krein skew-selfadjoint). For an even Lorentzian spectral triple, the $\mathbb Z_2$-grading $\chi$ must respect $\chi^*=\chi$, $\chi^2=1$, $[\chi,a] = 0$, $\chi
\J =- \J \chi$ and $\chi D =- D \chi $. The compactness condition on $a(1 +  \frac {(D D^* + D^*D)}2)^{-\frac 12}$ is not necessary for the metrical information but it is a natural generalization of the usual compact resolvent condition coming from the Riemannian case.

Then a Lorentzian distance, respecting the usual properties of non-negativity, antisymmetry and inverse triangle inequality, can be defined between two states $\varphi,\psi$ on $\A$ by:
\begin{equation}\label{ncgform}
 d(\varphi,\psi) \ =\ \inf_{a \in \A} \set{ \ [\psi(a)-\varphi(a)]^+ \ :\ \forall \phi \in \H, \scal{\phi,\J([D,a]+ i\chi) \phi } \leq 0\ },
   \ifiop
\else
\vspace{0.2cm}
\fi
 \end{equation}

where $i\chi$ should be replaced by $\pm 1$ (both signs) if the Lorentzian spectral triple is odd.

There exist two technical difficulties in applying the formula \eqref{ncgform} which we are going to discuss. However, we will show that in all currently existing examples, those difficulties can be bypassed.

The first difficulty concerns the fundamental symmetry $\J$. With the minimal set of axioms presented here concerning $\J$, there is no guarantee that the signature is exactly Lorentzian (it can correspond to a pseudo-Riemannian manifold in the commutative case) which means that the Lorentzian distance formula could give no result. The exact set of axioms in order to guarantee a Lorentzian signature is still an active subject of research, with some existing but not identical working proposals \cite{F5, Rennie12, BESNARD2017,CC2014}. Since we want to keep this paper as general as possible, we will not focus on one particular condition. The existing typical examples of noncommutative spacetimes, all of which could fit into the Lorentzian spectral triple formalism, are almost-commutative manifolds (Kaluza-Klein products of a usual Lorentzian manifold and a discrete noncommutative internal space \cite{SIGMA2014,JGP2015}) and deformations of flat spacetimes (Moyal spacetime, $\kappa$-Minkowski, Lorentzian cylinder, ... \cite{FrancoMoyal,cylinder}). For almost commutative manifolds, the suitable fundamental symmetry is $\J =\J_\M \otimes 1$ where $\J_\M$ is a fundamental symmetry for the based spacetime. For deformations of flat spacetimes, the canonical choice is $\J = i \gamma^0$. So, to our knowledge, there is no currently existing toy model (noncommutative noncompact complete Lorentzian spacetime) for which this problem must be taken into consideration, and the question is reserved for abstract considerations.

The second difficulty concerns the algebra $\A$ and the space of states on it. In traditional noncommutative geometry, $\A$ is a pre-C$^*$-algebra, since it corresponds in the commutative case to continuous functions vanishing at infinity. For causal considerations (causal functions), this algebra is too small and one must consider an additional specific unitization of $\A$ corresponding to bounded functions \cite{CQG2013}. However, steep functions present in the Lorentzian distance formula are clearly unbounded and cannot fit into the usual pre-C$^*$-algebra formalism. One must find a way to extend the initial pre-C$^*$-algebra corresponding to the states to unbounded elements and be sure that (some of) the states are still well-defined and uniquely extended. One particular way to realize such an extension is presented in \cite{F5}. Again, for all currently existing examples, the problem can be at least partially bypassed. For almost commutative manifolds, all pure states are product states between well-defined states on the based spacetime (evaluation maps) and vector states on the discrete algebra. For deformation spaces, the elements of the initial pre-C$^*$-algebra of bounded continuous functions are compact operators so all states correspond to vector states which can be easily and uniquely extended to unbounded functions as long as their evaluation remain finite, as used in \cite{FrancoMoyal}. Once more, this problem is an abstract one and does not prevent the application of the Lorentzian distance formula to particular models of noncommutative spacetimes.

\ifiop
\ack
\else
\vspace{0.5cm}
\section*{Acknowledgments}
\fi

The author would like to thank Ettore Minguzzi for organizing the meeting ``Non-Regular Spacetime Geometry'' between the two communities of Lorentzian geometry and noncommutative geometry, from which the evolution of the proof of the Lorentzian distance formula was possible.

\ifiop
\section*{References}
\bibliographystyle{iopart-num}
\bibliography{causality_bib}
\else
\vspace{0.5cm}
\setlength\bibitemsep{0.6\baselineskip}
\printbibliography
\fi
\end{document}